\documentclass[12pt]{article}
\pdfoutput=1
\usepackage{tikz}
\usetikzlibrary{matrix,arrows,decorations.pathmorphing,decorations.pathreplacing}
\usepackage{jheppub}
\usepackage{amsmath,amssymb,euscript,array,mathrsfs,appendix,ctable}
\usepackage{arydshln}
\usepackage{todonotes}
\usepackage{graphicx}

\def\BP{{\EuScript P}}
\def\BM{{\EuScript M}}

\def\BS{{\EuScript S}}
\def\BA{{\EuScript A}}

\def\BAP{{{\EuScript A}'}}

\def\BE{{\EuScript E}}

\def\BH{{\mathscr H}}

\newcommand{\Tr}{\operatorname{Tr}}

\newcommand{\RE}{\operatorname{Re}}

\def\Dbarslash{\,\,{\raise.15ex\hbox{/}\mkern-12mu {\bar D}}}
\def\Dslash{\,\,{\raise.15ex\hbox{/}\mkern-12mu D}}
\def\delslash{\,\,{\raise.15ex\hbox{/}\mkern-9mu \partial}}
\def\delbarslash{\,\,{\raise.15ex\hbox{/}\mkern-9mu {\bar\partial}}}

\def\bra#1{\langle #1|}
\def\ket#1{| #1\rangle}
\def\braket#1#2{\VEV{#1 | #2}}
\def\VEV#1{\langle #1\rangle}

\def\PP{\mathscr{P}}

\newcommand{\EQ}[1]{\begin{equation}\begin{split} #1
\end{split}\end{equation}}

\newcommand{\BOX}[1]{
\begin{center}\fbox{\parbox{15.8cm}{#1}}\end{center}}

\title{The Copenhagen Interpretation as an Emergent Phenomenon}
\author{Timothy J. Hollowood}
\affiliation{Particle Theory Group, Department of Physics, Swansea University, Swansea, SA2 8PP, U.K.}

\emailAdd{t.hollowood@swansea.ac.uk}

\abstract{The Copenhagen interpretation has been remarkably successful but seems at odds with the underlying linearity of quantum mechanics. We show how it can emerge in a simple way from the underlying microscopic quantum world governed by Schr\"odinger's equation without the need for observers or their brains. In order to achieve this, we assemble pieces of various pre-existing ideas. Firstly, we adopt a relational approach and use the eigenvectors of the reduced density matrix of a quantum sub-system, or equivalently the Schmidt decomposition, to define the ``internal state" of a sub-system. Previous work has identified serious objections to such an interpretation because it apparently leads to macroscopic superpositions and physically unacceptable instabilities near degeneracies. We show that both these problems are solved if the sub-system consists of a large number of coarse grained degrees of freedom as one expects in order to make contact with the classical world. We further argue that coarse graining is a necessary ingredient because measuring devices have both finite spatial and temporal resolutions. What results is an interpretation in which both decoherence and coarse graining play key roles and from which the rules of the Copenhagen interpretation are seen to emerge in realistic situations that include the measurement of the position of a particle and a decay process.
}

\notoc
\begin{document}

\maketitle

\newpage

\section{Introduction}\label{s1}

\pgfdeclarelayer{background layer} 
\pgfdeclarelayer{foreground layer} 
\pgfsetlayers{background layer,main,foreground layer}
\pgfdeclareimage[interpolate=true,width=11cm] {image5}{pic5}

The remarkable thing about questions involving the interpretation of quantum mechanics is that they seem to be irrelevant to the majority of physicists. This is because the seemingly ad-hoc rules of the Copenhagen interpretation as set out in most textbooks appear to be perfectly adequate for relating the rules of the theory to the phenomena that are observed.  

Nevertheless, at some point one wants to know if the Copenhagen interpretation can be derived from more fundamental laws. Given the success of the current state-of-affairs suggests that the new laws must of a minimal sort that do not change the theory in any really fundamental way. In particular, it seems desirable  that the quantum mechanics of an isolated system is described by the Schr\"odinger equation. However, at the same time one wants to understand why measurements on quantum system seem to give rise to distinct outcomes with probabilities that can be extracted in the standard way from the state vector. Basically one wants evolution by the Schr\"odinger equation in certain circumstances and something that appears to be a stochastic process in others. 
One way to unify these two different dynamical laws is to associate them to different views of the same system. 
We use the terminology of Bene \cite{Bene:1997mf}
and Sudbery \cite{Sud1} and refer to them as the ``external" and ``internal" views, respectively. For us the internal view is associated to a particular sub-system $\BA$ of a system $\BS$ and involves the reduced density matrix:

\BOX{
\noindent{\bf External view:} an isolated system $\BS$ will evolve according to the Schr\"odinger equation.

\noindent{\bf Internal view:}  the ``internal state" of a sub-system $\BA\subset\BS$ corresponds to one  eigenvectors of the reduced density matrix with a probability given by the corresponding eigenvalue.}

\noindent If we write $\BS=\BA+\BA'$, and suppose that the external state of $\BS$ is $\ket{\Psi}$, then the reduced density matrix has an eigen-decomposition of the form
\EQ{
\rho=\Tr_{\BH_{\BA'}}\ket{\Psi}\bra{\Psi}=\sum_{i=1}^N p_i\ket{\psi_i}\bra{\psi_i}\ ,\qquad
\braket{\psi_i}{\psi_j}=\delta_{ij}\ .
\label{red}
}
Note that $N$ is equal to the dimension of the smallest of the Hilbert spaces $\BH_\BA$ or $\BH_\BAP$ and the $\{p_i\}$ are a set of real numbers with $0\leq p_i\leq1$ and $\sum_ip_i=1$ and so can be given the interpretation of probabilities. For the moment we assume that there are no degeneracies in the decomposition, $p_i\neq p_j$ for $i\neq j$, but later we shall discuss degeneracies in detail. It is important that the complement $\BA'$ has a corresponding density matrix and eigen-decomposition
\EQ{
\rho'=\Tr_{\BH_\BA}\ket{\Psi}\bra{\Psi}=\sum_{i=1}^N p_i\ket{\psi'_i}\bra{\psi'_i}\ ,\qquad
\braket{\psi'_i}{\psi'_j}=\delta_{ij}\ ,
\label{red2}
}
with the same spectrum of eigenvalues $p_i$. The states $\ket{\psi_i}$ and $\ket{\psi'_i}$ are therefore correlated in a way that can be made manifest by describing the same situation in terms of the Schmidt decomposition \cite{Schm,Sch}. This is the unique decomposition (when $p_i\neq p_j$) of the total state vector $\ket{\Psi}$ in the form
\EQ{
\ket{\Psi}=\sum_{i=1}^N\sqrt{p_i}\ket{\Psi_i}\ ,\qquad\ket{\Psi_i}=\ket{\psi_i}\otimes\ket{\psi'_i}\ ,\label{sch}
}
with $\braket{\Psi_i}{\Psi_j}=\delta_{ij}$. Note that the reduced density matrices in \eqref{red} and \eqref{red2} follow immediately.

The idea of having different views of a quantum system is known as  a ``relational" or ``perspectival" interpretation \cite{Rov,Dk1,BD1}.  The idea of using the eigenvectors and eigenvalues of the reduced density matrix, or equivalently the Schmidt decomposition, in interpretations of quantum mechanics can be traced back to Zeh's work on decoherence \cite{Zeh70,Zeh73, Zeh00,Zeh05}. Using the Schmidt decomposition, or eigenvectors of the reduced density matrix to 
define states of a sub-system means that our interpretation is in the class of 
modal interpretations \cite{Barvinsky:1995va,Bene:1997mf,Bene:1997nd,Bene:1997kk,DV1,DV2,B1,Bu1,D1,D2,V1,Albrecht:1992rs,Albrecht:1992uc}.\footnote{See \cite{Kent:1996kx} for a discussion of the Schmidt decomposition and the consistent histories formulation of quantum mechanics.} As we develop the interpretation, we will find that 
the phenomenon of decoherence is 
of fundamental importance because it has a key bearing on the eigenvectors of the reduced density matrix. 
Decoherence itself has a large literature but some key papers are 
\cite{Zurek:1982ii,Caldeira:1982iu,Joos:1984uk,Zeh:1986ix,Unruh:1989dd,Zurek81,Zurek91,PHZ,ZHP,Zurek93,TS,ZP,GKZ,Gallis,AZ,APZ,Zurek98a,HSZ,Zurek98b,PZ2,Zeh99,BGJKS,Zurek03,KZ,Zeh10}
(the reviews \cite{JZKGKS,Schlosshauer:2003zy} contains many more references).
The goal of the present work is to take ideas from many of the works cited above and assemble them into a consistent interpretation of quantum mechanics that is based on clear and simple rules and that does not rely on the existence of observers and their brains.\footnote{We were particularly influenced by the works of Joos and Zeh \cite{Joos:1984uk}, Bene \cite{Bene:1997mf,Bene:1997nd,Bene:1997kk}, Bene and Dieks \cite{BD1}, Albrecht \cite{Albrecht:1992rs,Albrecht:1992uc} and Rovelli \cite{Rov}.} Alternative approaches will not be reviewed.

Many questions arise with interpretations based on the reduced density matrix but perhaps the most pressing is given a quantum system $\BS$ what determines the decomposition $\BA+\BA'$? There are obviously many ways to choose a sub-system $\BA$ each with their own internal view. The answer to this embarrassment of riches that we want to develop is that there is a distinguished sub-system $\BA$---or rather a class of closely related sub-systems.
These distinguished sub-systems correspond to coarse grained degrees of freedom
that match on to the classical world. The internal view of $\BA$ is necessarily, therefore, a coarse-grained one with all the microscopic degrees of freedom of the total system appearing in the complement $\BA'$. This still leaves open the question as to how to specify the 
sub-system $\BA$ in practice. In order to gain some insight, 
it is useful consider the same problem in the context of quantum field theory where there is a precise theory of coarse graining known as the renormalization group (for which Wilson won the Nobel Prize \cite{Wilson:1983}). In particular, in the quantum field theory context the conceptual framework is clear and what is true for quantum field theory must be true for quantum mechanics since the latter is just an approximation of the former.

\subsection{Lessons from quantum field theory}

The decomposition of a quantum system into the microscopic and coarse grained degrees of freedom
is very familiar in Quantum Field Theory (QFT). In this context, the notion of coarse graining is
a necessity. QFTs must be defined in the first instance with an explicit cut off on field modes in order to properly define the functional integral and regulate the UV divergences.
This can be in the form of a momentum cut-off $\mu$. The idea is that is we probe the theory on a momentum scales up to $p_0$, which we can view as the maximum scale of the resolution of our experiment, then we should take $\mu$ just above the scale $p_0$. This is because the theory with this cut off is the optimal description of the phenomena on momentum scales $<p_0$. So the theory with a cut off, the ``effective theory", has the relevant degrees of freedom at that scale. One may want to define a continuum theory by taking $\mu\to\infty$, however, this is only strictly relevant if one wants to ask physical questions that probe the theory at arbitrarily high momentum scales. In fact it is a very delicate issue whether the continuum limit $\mu\to\infty$ exists in any particular QFT.
For many theories, like quantum electrodynamics, the continuum limit very likely does not exist and as one probes the theory to higher momentum scales it actually breaks down (unless ``UV completed" by embedding in a larger theory like a grand unified gauge theory). This does not stop QED being remarkably successful for calculating low momentum phenomena. On the other hand, quantum chromodynamics QCD is completely different and does have a continuum limit due to the phenomenon of asymptotic freedom.

In the modern
way of thinking about QFT
the theory should only be defined as an effective theory with an in-build cut off scale that 
can float just above the scale set by experimental probes.
The process of changing of the cut off whilst keeping the physics on a low momentum scale $p_0$  fixed 
is the fundamental idea of the 
renormalization group. Cut offs can also be implemented in real space; for instance, the fields can be defined on a spatial lattice of spacing $\epsilon\sim\hbar/\mu$. Now the cut off $\epsilon$ is interpreted as a spatial resolution scale. Once again, if experiments only probe distance scales down to $x_0$ then the optimal choice for the cut off $\epsilon$ is just below the scale $x_0$. So there is a natural split of the field modes into those with momenta $>\mu$ (or spatial resolution $<\epsilon$), the ``microscopic" degrees of freedom, and modes with momenta $<\mu$ (with spatial resolution $>\epsilon$), the coarse-grained ``degrees of freedom". 

\BOX{The important conceptual point is that the resolution scale is set by the measurements that we perform; that is by the resolution of the measuring device.}

This modern way of thinking about QFTs has a message for quantum mechanics: although we usual think of the wave function $\psi(x)$ as function of a continuous spatial coordinate $x$ it is really only an effective quantity. Any description of the measurement process with a resolution scale $\epsilon$ will only depend on the wave function on this coarse grained scale. In QFT we usually implement the cut off at the level of the quantum field itself, but in quantum mechanics it is more convenient to build the cut off into the measuring apparatus rather than the wave function. Ultimately this is just a matter of taste and one could equally well define a regularized version of the wave function.
Whatever we do, the spatial cut off $\epsilon$, or spatial resolution of the measuring device, splits a quantum system into coarse grained and microscopic degrees of freedom which we can view as a decomposition of the Hilbert space\footnote{Balasubramanian, McDermott and Raamsdonk \cite{Balasubramanian:2011wt} investigate such decompositions of the Hilbert space in an interacting QFT and the extent to which there is entanglement between low and high  momentum modes when the QFT is in its vacuum state.}
\EQ{
\BH_\BS=\BH_\BA\otimes\BH_\BAP\ .
}
Note that the decomposition that we are advocating is very different from choosing the sub-system $\BA$ as the degrees of freedom in some spatial sub-region of the system $\BS$. For us it is crucial to have a split that involves the different length scales of the problem: microscopic versus coarse grained. To summarise:

\BOX{\noindent{\bf Finite resolution:}  the sub-systems of interest correspond to coarse grained degrees of freedom with some finite spatial resolution.}

The fact that measurement involves at some stage some kind of discreteness has been discussed previously by Joos and Zeh \cite{Joos:1984uk} and by Bene and Dieks \cite{BD1}. Note, however, that we view the discreteness as a fundamental aspect of a coarse grained description arising from the practical limitations of measuring devices. As a corollary,  the wave function as a smooth function is an idealization that may yield misleading results as we describe below and more fully in section \ref{s2}.

Returning to the issue of the identifying the internal view with an eigenvector of the reduced density matrix, there are two important problems that have been discussed in the literature:

\BOX{
\noindent{\bf The problem of non-locality of the eigenstates} \cite{B2,Donald1,Page:2011gi}: there is no guarantee that the  eigenstates $\ket{\psi_i}$ have reasonable classical properties. In fact in situations where the Hilbert space is spanned by a continuous spectrum of states the eigenvectors often involve a continuous superposition of macroscopically distinct states.

\noindent{\bf The problems of degeneracy} \cite{BDV,Donald1,Albrecht:1992uc}: when there are degeneracies amongst the eigenvalues $p_i$ it is not clear how to assign internal states. In addition, it has been shown that degeneracies give rise to other related problems. In some instances they lead to internal states that involve superpositions of macroscopically distinct states, in other instances the internal state can switch rapidly between macroscopically distinct states in a physically unacceptable way.}

All these problems must be solved if the interpretation being offered is to be convincing. As we will see, the degeneracy problem becomes bound up with how to define a dynamical law to the internal view when one recognises that the eigenvalues $p_i$ generally are time dependent as a consequence of the fact that the state $\ket{\Psi}$ of the total system satisfies the Schr\"odinger equation. Consistency of the external and internal views will require that the internal view is described by some stochastic process. This process is not determined uniquely but there appears to be a natural choice as we explain in section \ref{s4}.

\section{Decoherence, Finite Resolution and Localization}\label{s2}

In this section, we consider a simple set up consisting of a particle interacting with an environment which is modelled as a bath of very light particles originally discussed by Joos and Zeh \cite{Joos:1984uk}. In the first instance, we take $\BA$ to be the particle described by a wave function $\psi(x)$. The environment is then the complement $\BA'$. We will not need to describe the environment in any detail. The initial state of the system is some
non-entangled state $\ket{\Psi}=\ket{\psi}\otimes\ket{\psi'}$ and so the reduced density matrix of the particle in the position basis is simply
\EQ{
\rho(x,y)=\psi(x)\psi^*(y)\ .
}
The effect of the interaction with the environment has an effect on the reduced density matrix $\rho(x,y)$ that can be modelled by multiplying by a Gaussian factor \cite{Joos:1984uk}:
\EQ{
\rho(x,y)=\psi(x)\psi^*(y)e^{-(x-y)^2/\ell^2}\ ,
\label{xx1}
}
The coherence length $\ell$ has some specified time dependence depending on details of the model. In order to get a feel for the magnitudes involved, for the models of decoherence constructed by Joos and Zeh \cite{Joos:1984uk}, we can consider the case where the particle is a speck of dust of size $10^{-3}\ \text{cm}$ and the environment consists of air molecules. In this case, the coherence length behaves as
\EQ{
\ell\thicksim 10^{-18}t^{-1/2}\ \text{cm}\ .
}
So in $10^{-4}$ seconds the coherence length is of the order of the atomic scale.
Note that the model breaks down when $\ell$ reaches the thermal de~Broglie wavelength $\hbar/\sqrt{2\pi kT}$ of the particle.

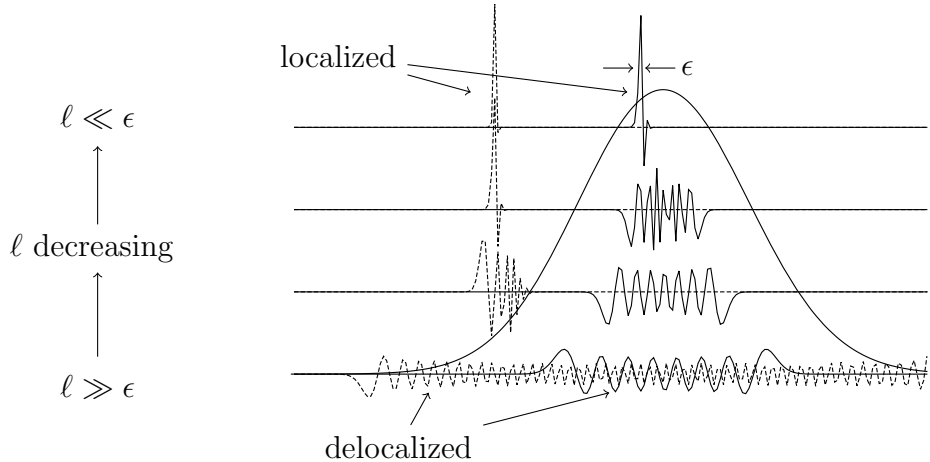
\begin{figure}
\begin{center}
\begin{tikzpicture}[scale=0.8]
\pgftext[at=\pgfpoint{0cm}{0cm},left,base]{\pgfuseimage{image5}} 
\node at (-3,0.3) (a1) {$\ell\gg\epsilon$};
\node at (-3,4.8) (a1) {$ \ell \ll\epsilon$};
\draw[->] (-3,0.8) -- (-3,2.2);
\draw[->] (-3,3) -- (-3,4.3);
\node at (-3.1,2.6) (a3) {$\ell$ decreasing};
\node at (2,-0.7) (b1) {delocalized};
\node at (1,5.8) (b2) {localized};
\draw[->] (b1) -- (2.5,0.2);
\draw[->] (b2) -- (3.2,5.2);
\draw[->] (b1) -- (5.5,0.2);
\draw[->] (b2) -- (5.8,5.2);
\draw[->] (5.4,5.6) -- (5.9,5.6);
\draw[<-] (6.1,5.6) -- (6.6,5.6);
\node at (6.8,5.6) (c1) {$\epsilon$};
\end{tikzpicture}
\caption{\small The figure shows the original wave function $\psi(x)$ (the Gaussian-like form) and the behaviour of two of the eigenvectors of the density matrix (continuous and dashed) from the regime where $\ell\gg\epsilon$ at the bottom to the regime $\ell\ll\epsilon$ at the top. The ensuing localization is very apparent as the coherence length $\ell$ drops below the spatial resolution scale $\epsilon$.}
\label{f1} 
\end{center}
\end{figure}

We now investigate the eigenvectors of the reduced density matrix in \eqref{xx1} for some choices $\psi(x)$. There are some particular choices that can be solved exactly which we present in appendix \ref{a1}, however, it is a simple matter to investigate what happens numerically. The simplest approach is to take a discretisation of \eqref{xx1} by defining it on a spatial lattice of size $\epsilon$ with points $x_j=j\epsilon$. As long as $\epsilon$ is smaller than the length scale implicit in $\psi(x)$ and the coherence length $\ell$, lattice effects are not important. One immediately finds that the eigenfunctions are not localised in any way, rather they are spread over the support of $\psi(x)$ as shown in figure \ref{f1}. So the eigenfunctions are in no way classical in the sense of being localised in space. This is perhaps somewhat surprising, one might have expected the Gaussian factor $\exp[-(x-y)^2/\ell^2]$ to localise the eigenfunction in position space on a scale set by $\ell$. But this intuition, as argued in \cite{B2,Donald1,Page:2011gi}, is flawed.

However, something interesting happens if we decrease the value of $\ell$ beyond the natural resolution of the system $\epsilon$ that we introduced by discretising the system. In that case, the eigenfunctions now become localized on the scale $\epsilon$ as illustrated in figure \ref{f1}.
One can also check that in the limit $\ell\ll\epsilon$, the eigenvalues $p_j$ approach the values 
$\epsilon|\psi(x_j)|^2$ as one would expect on the basis of the Copenhagen interpretation and Born's rule. So the trick that we used as a device to perform a numerical evaluation can be used as a form of coarse graining of the system with a finite resolution and this leads to internal states of the coarse grained sub-system, the particle, which are localised on the scale of the coarse graining and which have associated probability that are consistent with the Copenhagen interpretation.\footnote{This kind of lattice coarse graining is used widely in QFT. The spatial direction is taken to be discrete and the Hilbert space of the QFT is then generated by acting with creation operators $a_{x_j}^\dagger$, interpreted as creating a particle only at the discrete point $x_j=j\epsilon$, on the vacuum. The one-particle states $a_{x_j}^\dagger\ket{0}$ are then identified, in the non-relativistic limit, with the quantum mechanical states $\ket{x_j}$ and the wave function $\psi(x)$ is only defined at the lattice points with a general state being expressed as
$\ket{\psi}=\sum_j\psi(x_j)\ket{x_j}$.}

For the example of the dust particle in the model of Joo and Zeh, we expect that the resolution of a measuring device would be much larger than the atomic scale and so the localization time determined by $\ell(t)\sim\epsilon$, is smaller than $10^{-4}\ \text{s}$. For example, suppose $\epsilon\sim10^{-6}\ \text{cm}$ then the localization time is $10^{-6}\ \text{s}$.

\subsection{A simple model}

In this section we described in more detail a simple model of a particle interacting with a generic environment. At $t=0$ we suppose that the interaction is turned on and the initial state is the non-entangled state
\EQ{
\ket{\Psi(0)}=\ket{{\cal E}_0}\otimes\int_{x_1}^{x_{n+1}}dx\,\psi(x)\ket{x}\ .
}
For simplicity, we assume that $\psi(x)$ only has non-trivial support in the interval $x_1<x<x_{n+1}$. We suppose that the interaction Hamiltonian takes the form $H_\text{int}=H'\otimes\tilde x$, where $H'$ acts on the environment and $\tilde x$ is a finite resolution version of the position operator defined via
\EQ{
\tilde x\ket{x}=x_j\ket{x}\ ,\qquad x_j<x<x_{j+1}\ ,\qquad x_j=j\epsilon\ .
}
Solving the Schr\"odiner equation for $t>0$, gives
\EQ{
\ket{\Psi(t)}=\sum_{j=1}^n\ket{{\cal E}_j(t)}\otimes\int_{x_j}^{x_{j+1}} dx\,\psi(x)\ket{x}\ ,
\label{ww1}
}
where
\EQ{
\ket{{\cal E}_j(t)}=U_j(t)\ket{{\cal E}_0}\ .
}
Here, $U_j(t)$ is the effective unitary operator that implements time translations on the environment. Note that it depends implicitly on $j$ through the interaction Hamiltonian.

The reduced density matrix of the particle is
\EQ{
\rho=\sum_{ij=1}^n\Big(\sqrt{p_ip_j}\braket{{\cal E}_j(t)}{{\cal E}_i(t)}\Big)\ket{\psi_i}\bra{\psi_j}\ ,
\label{uu1}
}
where we have defined the orthonormal states of the particle
\EQ{
\ket{\psi_j}=\frac1{\sqrt{p_j}}\int_{x_j}^{x_{j+1}}dx\,\psi(x)\ket{x}\ , \qquad p_j=\int_{x_j}^{x_{j+1}}dx\,|\psi(x)|^2\ .
\label{bb1}
}

The environment consists of a large number of components which are set in motion by the interaction with the particle. Since the effective Hamiltonian $H_j$ depends on $j$ we expect that the inner products $\braket{{\cal E}_i(t)}{{\cal E}_j(t)}$, for $i\neq j$, will become small. In appendix \ref{a3} we discuss a model of the environment consisting of large number of spin $\frac12$ systems coupled to the particle. In this case, we find
\EQ{
\braket{{\cal E}_i(t)}{{\cal E}_j(t)}\thicksim \exp\big[-N\lambda^2 t^2(x_i-x_j)^2\big]=\exp\big[-N(\lambda\epsilon t)^2(i-j)^2\big]\ .
}
where $N$ is the number of spins involved and $\lambda$ is an average coupling. This kind of exponential fall-off is expected to be generic. With this behaviour, the eigenvectors of $\rho$ in \eqref{uu1}, will quickly approach the states $\ket{\psi_j}$ with eigenvalues $p_j$. So for $t\gg N^{-1/2}(\lambda\epsilon)^{-1}$, the internal states of the particle 
are approximately $\ket{\psi_j}$ with probabilities $p_j$. By approximate, we mean up to terms of the form $\exp[-N(\lambda\epsilon t)^2]$ which even for only a modestly large, but still sub-macroscopic, environment, say $N\sim 10^{10}$, will be extraordinarily small due to the double exponential. The internal state of the particle therefore has a well defined position up to the level of the resolution scale $\epsilon$. The fact that a large number of components leads to 
a huge suppression of the inner products of macroscopically distinct states has been stressed by Banks \cite{Banks:2009gw}.  

\section{Measurement}\label{s3}

One might argue that we introduced the coarse graining in the last section by hand. However, the philosophy we are adopting here is that the cut off $\epsilon$ is set to the scale of the resolution of the measurements that are performed on the particle. One way to implement this philosophy is to 
actually include an explicit measuring device $\BM$ with some realistic in-built spatial resolution $\epsilon$ in the description. The philosophy could be paraphrased by saying that a particle is what a measuring device detects, a viewpoint that is essential when one formulates QFT in a curved space-time, but has a wider applicability. 
Attention can then switch from the particle itself to its effect on the measuring device. In this situation the measuring device itself provides the necessary decoherence and, at least to start with, a separate environment need not be considered. We will now develop this model in some detail.

A simple model with a finite resolution is obtained by taking the particle environment system in the last section and re-interpreting the environment as the measuring device $\BM$. In this re-interpretation the states $\ket{\BE_j(t)}\to\ket{\BM_j(t)}$, for large enough $t$, need to be macroscopically distinct states that indicate whether the particle was measured in the spatial interval $x_j<x<x_{j+1}$, with $x_j=j\epsilon$, where $\epsilon$ is the spatial resolution of $\BM$. In this simple model, we are ignoring all the other microscopic degrees of freedom---the environment---but later we will find the it plays a crucial r\^ole when we consider measuring inefficiencies and problems involving degeneracy.

Taking the external state as in \eqref{ww1} with $\ket{\BE_j(t)}\to\ket{\BM_j(t)}$, the reduced density matrix of $\BM$ takes the form
\EQ{
\rho=\sum_jp_j\ket{\BM_j(t)}\bra{\BM_j(t)}\ .
\label{rdm}
}
At $t=0$ the measuring device starts in some initial resting state $\ket{\BM_0}$ (with unit probability). For $t>0$ \eqref{rdm}
is not an eigenvector decomposition of $\rho$ because the states $\ket{\BM_j(t)}$ are not orthogonal. However, the important point is that for large enough $t$ the states $\ket{\BM_j(t)}$ become macroscopically distinct and so will have a minuscule inner product, just like the environment states $\ket{\BE_j(t)}$ discussed previously. 
In order to make this point, let us estimate the inner product. Suppose that $\BM$ is modelled by $N$ coarse-grained constituents
whose individual wave functions are localized in space to the resolution scale $\epsilon$. Then roughly speaking, if the difference between the states at some time $t$ involves a macroscopic distance scale $X$, then the inner product is 
\EQ{
\big|\braket{\BM_i(t)}{\BM_j(t)}\big|\thicksim \exp\big[-NX^2/\epsilon^2\big]\ .
\label{ipn}
}
Even if we are overly generous and assume that the scale of the coarse graining is quite large say $\epsilon\sim10^{-2}\ \text{cm}$ and $N$ is quite small, say $10^3$, with $X\sim 1\ \text{cm}$ we have an inner product $\sim\exp(-10^7)$ which is an incredibly small number.
Therefore for large enough $t$, the state $\ket{\BM_j(t)}$ will be an approximate eigenstate of $\rho$ up to very small admixtures of the other states $\ket{\BM_i(t)}$, $i\neq j$, with an eigenvalue that is approximately $p_j$. Let us emphasize that that fact that $\BM$ has a finite resolution and the set of states is discrete is crucial for this conclusion. Therefore, to a very good approximation, after sufficient time:

\BOX{{\bf The internal state:}  is one of the macroscopically distinct states $\ket{\BM_j(t)}$ with probability $p_j=\int_{x_j}^{x_{j+1}}dx\,|\psi(x)|^2$ in accordance with the Copenhagen interpretation and Born's rule.}
\begin{figure}[ht]
\begin{center}
\begin{tikzpicture}[yscale=0.2,xscale=0.3]
\draw[->] (0,0) -- (20,0);
\draw[->] (0,0) -- (0,21);
\node at (21,0) (a1) {$t$};
\node at (-1.2,19) (a1) {$1$};
\node at (-1.2,0) (a1) {$0$};
\node at (26,13) (b1) {$\ket{\BM_1(t)}$};
\node at (26,6) (b2) {$\ket{\BM_2(t)}$};
\node at (26,1) (b3) {$\ket{\BM_3(t)}$};
\node at (-3,22.5) (b4) {$\ket{\BM_0}$};
\draw[->] (b4) -- (-0.2,19);
\draw[->] (b1) -- (19.2,11.4);
\draw[->] (b2) -- (19.2,5.7);
\draw[->] (b3) -- (19.2,1.9);
\node[rotate=90] at (-2,10) (a1) {probability};
\draw[decoration={brace,amplitude=0.5em},decorate] (8,20) -- (19,20);
\draw[decoration={brace,amplitude=0.5em},decorate] (0.4,20) -- (6.5,20);
\node at (3.3,22.5) (z1) {collapse region};
\node at (13.5,26.5) (a1) {$\ket{\BM_j(t)}$};
\node at (13.5,24.5) (a1) {macroscopically};
\node at (13.5,22.5) (a1) {distinct};
\draw[-] (6.5,0) -- (6.5,-0.5);
\node at (6.5,-2) (d1) {$t_\text{c}$}; 
\draw[very thick] plot[smooth] coordinates {(0, 19.)  (1, 18.2124)  (2, 16.4584)  (3, 14.6642)  (4, 13.2187)  (5,12.2299)  (6, 11.6976)  (7, 11.4838)  (8, 11.4189)  (9, 
  11.4035)  (10, 11.4005)  (11, 11.4001)  (12, 11.4)  (13, 11.4)  (14,
   11.4)  (15, 11.4)  (16, 11.4)  (17, 11.4)  (18, 11.4)  (19, 11.4)};
\draw[very thick] plot[smooth] coordinates {(0, 0)  (1, 0.773204)  (2, 2.35062)  (3, 3.65956)  (4, 
  4.51829)  (5, 5.12001)  (6, 5.48233)  (7, 5.6376)  (8, 5.68586)  (9,
   5.69741)  (10, 5.69961)  (11, 5.69995)  (12, 5.7)  (13, 5.7)  (14, 
  5.7)  (15, 5.7)  (16, 5.7)  (17, 5.7)  (18, 5.7)  (19, 5.7)};
\draw[very thick] plot[smooth] coordinates  {(0, 0)  (1, 0.0143922)  (2, 0.191004)  (3, 
  0.676266)  (4, 1.26304)  (5, 1.65009)  (6, 1.82005)  (7, 
  1.87861)  (8, 1.89526)  (9, 1.89913)  (10, 1.89987)  (11, 
  1.89998)  (12,1.9)  (13,1.9)  (14,1.9)  (15,1.9)  (16,1.9)  (17, 1.9)  (18, 1.9)  (19, 1.9)};
\end{tikzpicture}
    \end{center}
  \caption{\small The probabilities as a function of time for the reduced density matrix \eqref{rdm} in the case of three distinct measurement results. When the states $\ket{\BM_i(t)}$ become macroscopically distinct the inner products $\braket{\BM_i(t)}{\BM_j(t)}$ are
minuscule and the eigenstates of the reduced density matrix are approximately the states
$\ket{\BM_j(t)}$ with probabilities $p_j$.}
\label{f5}
\end{figure}
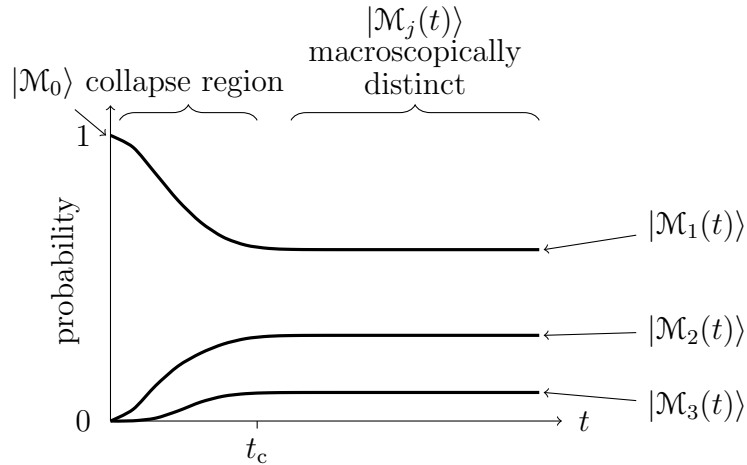
The behaviour of the probabilities is illustrated in figure \ref{f5}. Notice that there is an initial transient region when the probabilities are changing in time whose interpretation we shall address in section \ref{s4}. This is the region that the Copenhagen interpretation would identify as where the collapse occurs.
We can estimate the time of the collapse as follows.
If the measurement process takes a time $T$ and given \eqref{ipn}, the 
collapse time is roughly
\EQ{
t_\text{c}\thicksim \frac{T\epsilon}{XN^{1/2}}\ .
\label{loc}
}
After this time, the internal states are fixed: in the language of the Copenhagen interpretation the wave function has collapsed. This means that the probabilities are only changing significantly for $t<t_\text{c}$ when the states $\ket{\BM_j(t)}$ are not yet macroscopically distinguishable.  The results here are identical to the behaviour seen in the models developed by Albrecht \cite{Albrecht:1992rs,Albrecht:1992uc}. In those models the system being measured is the spin of a spin $\frac12$ particle and the interaction is modelled by random Hamiltonians in a high dimensional Hilbert space.

Note if we include the environment in the discussion so that the total state of the system is
\EQ{
\ket{\Psi(t)}=\sum_{j=1}^n\sqrt{p_j}\ket{\BM_j(t)}\otimes\ket{\psi_j}\otimes\ket{{\cal E}_j(t)}\ ,
\label{ww2}
}
does not change the conclusion above because the reduced density of the sub-system $\BM$ is still \eqref{rdm} as a result of the states $\ket{\psi_j}$ being orthogonal.

\subsection{Problems with degeneracy}\label{s3.1}

There are several facets to the problem with degeneracy.
The most basic one is if the density matrix $\rho$ of $\BA$ has degenerate eigenvectors then what states in the degenerate subspace correspond to the internal states of $\BA$? One way to solve this problem is to argue that the eigenstates and probabilities 
are analytic functions of $t$ and so one can assert that the internal states correspond to the analytic continuation of the eigenstates to the point of degeneracy (this is discussed in \cite{BDV}). For the moment this looks adequate, but later, in section \ref{s4}, we will show that actually it is not a degeneracy that causes a real problem but a near degeneracy.

The second issue concerning degeneracies appears at first sight to be more serious and has been discussed at some length by Albrecht \cite{Albrecht:1992uc}.
In order to make the point as simply as possible, let us consider a very coarse measurement of the position of a particle which only has two possible outcomes corresponding to $x_1<x<x_2$ and $x_2<x<x_3$.\footnote{Alternatively one could consider the measurement of the spin of a spin $\frac12$ particle.} If the initial state of the particle is such that 
$p_1=\frac12+\delta$ and $p_2=\frac12-\delta$ then the reduced density matrix of $\BM$ 
takes the form in \eqref{rdm}. In the $\ket{\BM_j(t)}$ basis
\EQ{
\rho_{ij}=\bra{\BM_i(t)}\rho\ket{\BM_j(t)}=\left(\begin{array}{cc}\frac12+\delta & \Delta(t)\\ \Delta(t)^*&\frac12-\delta\end{array}\right)\ ,
\label{tvv}
}
where 
\EQ{
\Delta(t)=\braket{\BM_1(t)}{\BM_2(t)}\ .
}
As long as $\delta$ is finite, then for $|\Delta(t)|\ll\delta$, the eigenvectors of $\rho$ are approximately equal to the two states $\ket{\BM_j(t)}$ with eigenvalues $\frac12\pm\delta$, respectively.\footnote{Note that the use of approximately here means up to terms of order \eqref{ipn}.}
Physically this seems perfectly acceptable since $|\Delta(t)|$ is so small.
However, the problem arises when $\delta$ vanishes identically, since then for $|\Delta(t)|\ll1$ the eigenvectors correspond to to the macroscopic superpositions
\EQ{
\ket{\BM_1(t)}\pm\ket{\BM_2(t)}\ ,
\label{vsa}
}
which would be disastrous for the interpretation.

One reaction to this is to say that the fine tuning required to set $\delta=0$ exactly is impossible to arrange in practice; however, this seems rather an unsatisfactory response; indeed, in other 
situations we can use symmetry to enforce the fine tuning. For example, consider measuring the spin along some direction of a spin $\frac12$ particle in a two particle singlet state. In this case rotational symmetry ensures that the probabilities of spin up and spin down are both $\frac12$ exactly.
A better argument is to acknowledge that our treatment of the measuring process has been too simple and, in particular, we should reconsider the effect of the environment $\BE$ and also measuring inefficiencies. Note that the environment consists of all the microscopic degrees of freedom of the measuring device along with what would identify 
as the external environment. In general, therefore, we have a tripartite decomposition of the total system $\BS=\BM+\BP+\BE$, corresponding to the coarse grained degrees of freedom of the measuring device, the particle and the environment. What is relevant is the internal view of $\BM$ obtained by tracing out the states in $\BP+\BE$. We now argue that the effect of including the environment is to split each outcome that originally had a probability $p_j$ into a finer set with probabilities that sum to $p_j$. The finer set of probabilities will inevitably be at least weakly time dependent and so
the problem identified above leading to a macroscopic superposition \eqref{vsa} never occurs. 
 
Including the environment, a more accurate description of the measurement process involves a state of the form
\EQ{
\ket{\Psi(t)}=\sum_{j=1}^n\sqrt{p_j}\sum_{a=1}^{m_j}Z_{aj}(t)\ket{\BM_{aj}(t)}\otimes\ket{\psi_j(x)}\otimes\ket{\BE_{aj}(t)}\ ,
\label{vv5}
}
where, for large $t$, $\ket{\BM_{aj}(t)}$ for fixed $j$ are a set of states clustered around a particular macroscopically distinct state of $\BM$ indicating that the particle has been detected in the interval $x_j<x<x_{j+1}$. In this description, the states $\ket{\BM_{aj}(t)}$, for a given $j$, are not necessarily themselves macroscopically distinguishable. The weighting factors $Z_{aj}(t)$ must satisfy
\EQ{
\sum_{a,b=1}^{m_j}Z_{aj}(t)^*Z_{bj}(t)\braket{\BM_{aj}(t)}{\BM_{bj}(t)}\braket{\BE_{aj}(t)}{\BE_{bj}(t)}=1\ ,
}
for each $j$, in order that $\ket{\Psi(t)}$ is normalized (as it must be since according to the external view $\ket{\Psi(t)}$ solves the Schr\"odingier equation).

The reduced density matrix of $\BM$ now has the form
\EQ{
\rho(t)=\sum_{j=1}^n\sum_{a,b=1}^{m_j}\Big(p_jZ_{aj}(t)^*Z_{bj}(t)\braket{\BE_{aj}(t)}{\BE_{bj}(t)}\Big)\ket{\BM_{aj}(t)}\bra{\BM_{bj}(t)}\ .
}
For large enough $t$, the states of the environment $\ket{\BE_{aj}(t)}$ will become orthogonal in both their indices up to the usual hugely suppressed Gaussian factors. 
The coarse-grained states of the measuring device $\ket{\BM_{aj}(t)}$ are expected to be approximately orthogonal in their $i,j$ indices, as in \eqref{ipn}, but not necessarily in their $a,b$ indices. This is not a problem since the states are only macroscopically distinct for $i\neq j$. Hence, to a high degree of accuracy 
\EQ{
\rho(t)=\sum_{j=1}^n\sum_{a=1}^{m_j}p_j|Z_{aj}(t)|^2\ket{\BM_{aj}(t)}\bra{\BM_{aj}(t)}\ .
}
Furthermore, since the states $\ket{\BM_{aj}(t)}$ are to a high degree of accuracy orthogonal for $i\neq j$, the eigenvalue problem for $\rho$ decomposes into $n$ subspaces labelled by $j$:
\EQ{
\rho&=\sum_{j=1}^n\rho^{(j)}\ ,\quad
\rho^{(j)}(t)=p_jR^{(j)}(t)\ ,\\ R^{(j)}(t)&=\sum_{a=1}^{m_j}|Z_{aj}(t)|^2\ket{\BM_{aj}(t)}\bra{\BM_{aj}(t)}\ ,
}
so the internal states of $\BM$ correspond approximately to the 
eigenvectors of $\rho^{(j)}$, for each $j$, with eigenvalues $p_j\xi^{(j)}_a(t)$ where $\xi^{(j)}_a(t)$ is the corresponding eigenvalue of $R^{(j)}(t)$.
Notice that  the original set of probabilities $\{p_j\}$ has now been split into a finer set $\{p_j\xi^{(j)}_{a}(t)\}$. However, the overall probability that measurement yields the result that the particle is in the interval $x_j<x<x_{j+1}$ is 
\EQ{
\text{Tr}\,(\rho^{(j)})=p_j\text{Tr}\,(R^{(j)})=p_j\ ,
}
as before. The important point is that the issue of whether there is a degeneracy $p_i=p_j$ now becomes irrelevant. One might argue that there is now a new degeneracy problem whenever $p_i\xi^{(i)}_b(t)=p_j\xi_a^{(j)}(t)$ which could occur at some particular instant of time. However, as we have mentioned previously, this kind of instantaneous degeneracy is not a problem because one can rely on the analyticity of the eigenvectors to define the internal state at the degenerate point and these states do not involve macroscopic superpositions.
Unfortunately, this does not completely solve the problem of degeneracy because a new problem arises when two states become almost degenerate, as we explain in section \ref{s4.2}.

Another facet of measurement is that no measuring device can be 100\% efficient and there will always be some chance that a result $\ket{\BM_i(t)}$ will be indicated even though the particle is in the state $\ket{\psi_j}$, $j\neq i$. 
Albert and Loewer \cite{AL1} originally suggested that such inefficiencies would lead to internal states involving macroscopic superpositions. However, their analysis failed to include the environment and once its effects are included, as originally pointed out by Bacciagaluppi and Hemmo \cite{BH1},  physically satisfactory results are obtained with no macroscopic superpositions. For completeness, measurement inefficiencies are discussed in appendix \ref{a2}.

\section{Time Dependence}\label{s4}

As it stands the eigenvectors of the reduced density matrix $\ket{\psi_i}$ determine the possible internal states of $\BA$ and their probabilities $p_i$ at a given time. 
In the collapse region, where the probabilities are time dependent, there must inevitably be transitions between the internal states. One might think that in a coarse grained description
this transient region of time dependence is too fleeting to be resolved and so one could simply assert that it is only the final probabilities that matter in practice. However, there are other kinds of measurement scenarios, like the one described in section \ref{s5}, where the time dependence occurs on macroscopic temporal scales and so an investigation of time dependence cannot be avoided.

Following our general philosophy, we will seek 
a description of the dynamics which is coarse grained. The simplest way of implementing this
is to describe the system at some discrete time
steps $t_n=n\eta$, $n=0,1,2,\ldots$. A ``history" is then sequence of internal states 
\EQ{
\Big\{\ket{\psi^{(0)}_{i_0}},\ket{\psi^{(1)}_{i_1}},\ket{\psi^{(2)}_{i_2}},\ldots\Big\}\ .
\label{plk}
}
The most natural kind of stochastic process to describe the histories is a Markov chain in which the probabilities at time step $t_{n+1}$ are related to those at step $t_n$ via a relation of the form
\EQ{
p_i^{(n+1)}=\sum_jp_{ij}^{(n)}p_j^{(n)}\ ,
\label{vv1}
}
with $\sum_ip_{ij}^{(n)}=1$. Notice that in our situation, the single time probabilities $p_i^{(n)}$ are given as the eigenvalues of the reduced density matrix at time $t_n$, 
so that the conditional probabilities $p_{ij}^{(n)}$ are thereby constrained. However, they are not determined uniquely by \eqref{vv1} and any specification of the $p_{ij}^{(n)}$ must be viewed as an additional hypothesis. The question is whether there is a natural way to define the conditional probabilities that seems physically reasonable and then whether the hypothesis leads to consequences that can be tested. There does indeed turn out to be is a natural choice and furthermore there are types of measurement that support it; this latter point being explored in section \ref{s5}. 

The natural choice, in a sense to be explained, is motivated by first studying the 
continuous time process obtained in the limit $\eta\to0$, although ultimately we will argue that 
one must not take $\eta\to0$ and this coarse graining is an essential step in getting a physically acceptable stochastic process. A careful discussion of the possible stochastic processes involved in given in \cite{Bacciagaluppi:1997da}.
Consider the Schmidt decompositon \eqref{sch}.
Given the state of the complete system $\ket{\Psi}$ satisfies the Schr\"odinger equation, it follows that
\EQ{
\frac{dp_i}{dt}=2\sqrt{p_i}\,\RE\,\bra{\Psi}\Big(\frac{\partial}{\partial t}+\frac i\hbar H\Big)\ket{\Psi_i}\ ,
\label{vv3}
}
where $H$ is the Hamiltonian of the complete system. 
Notice that 
this expression manifests the fact that if $\ket{\Psi_i}$ actually solves the Schr\"odinger equation, which would be the case
if there were no interactions between the two sub-systems, $H=H_\BA+H_{\BA'}$, then the probabilities
$p_i$ would be constant.
From \eqref{vv3}, we can then write
\EQ{
\frac{dp_i}{dt}=\sum_jJ_{ij}\ ,\qquad J_{ij}=-J_{ji}\ ,
\label{ge1}
}
with
\EQ{
J_{ij}=2\sqrt{p_ip_j}\,\RE\,\bra{\Psi_j}\Big(\frac{\partial}{\partial t}+\frac i\hbar H\Big)\ket{\Psi_i}\ .
\label{eqj}
}

What is nice about \eqref{eqj} is that the terms $J_{ij}$ just depend on the two states $\ket{\Psi_i}$ and $\ket{\Psi_j}$. The form \eqref{ge1} suggests an interpretation as a continuous Markov process in which there is a rate $T_{ij}$ that the state $\ket{\Psi_j}$ makes a transition to the state $\ket{\Psi_i}$. This would give
\EQ{
\frac{dp_i}{dt}=\sum_{j\neq i}\Big(T_{ij}p_j-T_{ji}p_i\Big)\ ,
\label{ge2}
}
corresponding to transitions in and out of  $\ket{\Psi_i}$. Although \eqref{ge1} and \eqref{ge2} have a similar form there is
no unique solution for the transition rates $T_{ij}$ in terms of the $J_{ij}$. 
Physically, though, it is natural that the transition rate $T_{ij}$ should only depend on the states $\ket{\Psi_i}$ and $\ket{\Psi_j}$ and furthermore
transitions between macroscopically distinct states should be highly suppressed. This latter point is subtle, as we explain in the next section. However, the natural choice is to take the transition rates to be one way meaning that if $T_{ij}\neq0$ then $T_{ji}=0$, and vice-versa. In this case one has 
\EQ{
T_{ij}=\frac1{p_j}\,\text{max}\left(J_{ij},0\right)\ .
\label{trt}
}
This is the choice made by Bell \cite{Bell} and discussed by Bacciagaluppi and Dickson \cite{Bacciagaluppi:1997da} and Sudbery \cite{Sud1}.

\begin{figure}[ht]
\begin{center}
\begin{tikzpicture}[yscale=0.2,xscale=0.3]
\draw[->] (0,0) -- (20,0);
\draw[->] (0,0) -- (0,21);
\node at (21,0) (a1) {$t$};
\node at (-1.2,19) (a1) {$1$};
\node at (-1.2,0) (a1) {$0$};
\node[rotate=90] at (-2,10) (a1) {probability};
\draw[-] (6.5,0) -- (6.5,-0.5);
\node at (6.5,-2) (d1) {$t_\text{c}$}; 
\draw[very thick] plot[smooth] coordinates {(0, 19.)  (1, 18.2124)  (2, 16.4584)};
\draw[->,very thick]  (2, 16.4584) --  (2, 2.35062);
\draw[densely dashed] plot[smooth] coordinates {(2, 16.4584)  (3, 14.6642)  (4, 13.2187)  (5,
   12.2299)  (6, 11.6976)  (7, 11.4838)  (8, 11.4189)  (9, 
  11.4035)  (10, 11.4005)  (11, 11.4001)  (12, 11.4)  (13, 11.4)  (14,
   11.4)  (15, 11.4)  (16, 11.4)  (17, 11.4)  (18, 11.4)  (19, 11.4)};
\draw[densely dashed] plot[smooth] coordinates {(0, 0)  (1, 0.773204)  (2, 2.35062) };
\draw[very thick] plot[smooth] coordinates { (2, 2.35062)  (3, 3.65956) (4,4.51829)  (5, 5.12001) };
\draw[->,very thick]  (5, 5.12001)  --   (5, 1.65009) ;
\draw[densely dashed] plot[smooth] coordinates
  {(5, 5.12001)  (6, 5.48233)  (7, 5.6376)  (8, 5.68586)  (9,
   5.69741)  (10, 5.69961)  (11, 5.69995)  (12, 5.7)  (13, 5.7)  (14, 
  5.7)  (15, 5.7)  (16, 5.7)  (17, 5.7)  (18, 5.7)  (19, 5.7)};
\draw[densely dashed] plot[smooth] coordinates  {(0, 0)  (1, 0.0143922)  (2, 0.191004)  (3, 
  0.676266)  (4, 1.26304) (5, 1.65009)};
 \draw[very thick] plot[smooth] coordinates  {  (5, 1.65009)  (6, 1.82005)  (7, 
  1.87861)  (8, 1.89526)  (9, 1.89913)  (10, 1.89987)  (11, 
  1.89998)  (12, 1.9)  (13, 1.9)  (14, 1.9)  (15, 1.9)  (16, 
  1.9)  (17, 1.9)  (18, 1.9)  (19, 1.9)};
\end{tikzpicture}
\end{center}
\caption{\small Following on from figure \ref{f5}, an example of a history of an internal state of $\BM$ according to the stochastic process. Note that transitions only occur between the internal states when they are not yet macroscopically distinct in the collapse region $t<t_\text{c}$. Once the states have become macroscopically distinct then transitions are highly suppressed and the internal states are equal to $\ket{\BM_j(t)}$ to a high degree of accuracy.}
\label{f6}
\end{figure}
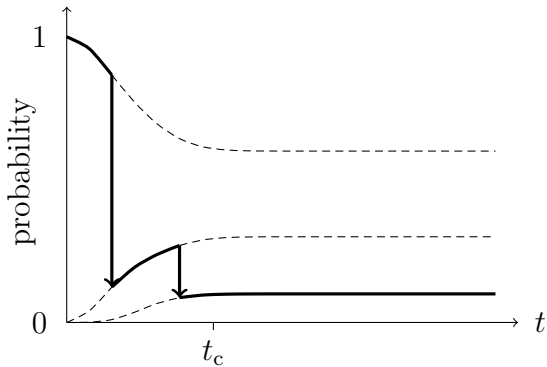

The fact that the transition rate $T_{ij}$ is determined by a matrix element involving the states  $\ket{\Psi_i}$ and $\ket{\Psi_j}$ is crucial to get a physically sensible theory. It ensures that transitions between macroscopically distinct states are highly suppressed. Transitions only occur when probabilities are changing in time as illustrated in figure \ref{f6} when the states are not yet macroscopically distinct. In this region the internal states do not satisfy Schr\"odinger equation.
After the initial transient period $t>t_\text{c}$, the internal states, that are to a very close approximation the states $\ket{\BM_j(t)}$, will then satisfy the Schr\"odinger equation and the probabilities are constant. So we see that key features of the Copenhagen interpretation are reproduced and the collapse of state vector becomes a physical process according to the internal view. The situation is illustrated in figure \ref{f6} which show transitions in the initial transient period. On the contrary, according to the external view, the Schr\"odinger equation is always satisfied. 

\subsection{Another problem with degeneracy}\label{s4.2}

Unfortunately the continuous time stochastic process described above has a fundamental flaw as pointed out in \cite{BDV,Donald1}. The problem concerns the behaviour of the system when two of the probabilities, say $p_i(t)$ and $p_j(t)$ become nearly degenerate. Consider the measurement system described in section \ref{s3} and focus on two particular components of the state corresponding to $\ket{\BM_i(t)}$ and $\ket{\BM_j(t)}$ which are approximate eigenstates of $\rho$ with eigenvalues that become degenerate at, say, $t=t_0$.\footnote{In order that the eigenvalues are time dependent requires the more sophisticated model including the environment in which case the states are labelled with pairs of indices $aj$. For simplicity we will just use a single label $j$ here.} In order to make the argument we need the two states to be already macroscopically distinct. Isolating the important part of the reduced density matrix in the 2-dimensional subspace, near the degenerate point we have 
\EQ{
\rho_\text{subspace}\thicksim\left(\begin{array}{cc} p_0+a(t-t_0)& p_0\Delta\\ p_0\Delta^*& p_0-a(t-t_0)\end{array}\right)\ ,
}
where \eqref{ipn}
\EQ{
\Delta=\braket{\BM_i(t)}{\BM_j(t)}\thicksim\exp\big[-NX^2/\epsilon^2\big]\ ,
}
is minuscule (near the degeneracy point we will ignore the time dependence of $\Delta$).
A true degeneracy requires both $t=t_0$ and $\Delta=0$ and consequently has co-dimension 3 in the space of density matrices. This is actually a general result which proves that a near degeneracy is actually the generic situation \cite{BDV,Donald1}. Suppose at $t=t_0$,  we have the non-generic situation where $\Delta=0$ exactly so that the system actually passes through the degenerate point. In this case the eigenvectors of $\rho$ for $t\neq 0$ are approximately equal to $\ket{\BM_i(t)}$ and $\ket{\BM_j(t)}$, up to order $\Delta$. We can then define the internal states for all $t$ including $t=t_0$ by using the analyticity of the eigenvectors in $t$. Note that nothing discontinuous happens at the degenerate point.

\begin{figure}[ht]
  \begin{minipage}[t]{.47\textwidth}
    \begin{center}
\begin{tikzpicture}[scale=0.2]
\node at (-13,10) (a1) {$\ket{\BM_i}$};
\node at (-13,-10) (a1) {$\ket{\BM_j}$};
\node at (13,10) (a1) {$\ket{\BM_j}$};
\node at (13,-10) (a1) {$\ket{\BM_i}$};
\node at (10.5,0) (a2) {$2|p_0\Delta|$};
\node at (0,-8) (a2) {$|p_0\Delta/a|$};
\draw[->] (-17,0) -- (-13,0);
\draw[->] (-17,0) -- (-17,4);
\node at (-11,0) (c1) {$t$};
\node at (-17,6) (c2) {prob};
\draw[decoration={brace,amplitude=0.5em},decorate] (6,2) -- (6,-2);
\draw[decoration={brace,amplitude=0.5em},decorate] (3,-5) -- (-3,-5);
\draw[very thick] plot[smooth] coordinates {(-10., 10.1489)  (-9., 9.16515)  (-8., 8.18535)  (-7., 7.2111)  (-6.,
   6.245)  (-5., 5.2915)  (-4., 4.3589)  (-3., 3.4641)  (-2., 
  2.64575)  (-1., 2.)  (0., 1.73205)  (1., 2.)  (2., 2.64575)  (3., 
  3.4641)  (4., 4.3589)  (5., 5.2915)  (6., 6.245)  (7., 7.2111)  (8.,
   8.18535)  (9., 9.16515)  (10., 10.1489)};
\draw[very thick] plot[smooth] coordinates {(-10., -10.1489)  (-9., -9.16515)  (-8., -8.18535)  (-7., -7.2111) 
(-6., -6.245)  (-5., -5.2915)  (-4., -4.3589)  (-3., -3.4641)  (-2., 
-2.64575)  (-1., -2.)  (0., -1.73205)  (1., -2.)  (2., -2.64575)  
(3., -3.4641)  (4., -4.3589)  (5., -5.2915)  (6., -6.245)  (7.,
-7.2111)  (8., -8.18535)  (9., -9.16515)  (10., -10.1489)};   
\end{tikzpicture}
    \end{center}
  \end{minipage}
  \hfill
  \begin{minipage}[t]{.47\textwidth}
    \begin{center}
\begin{tikzpicture}[xscale=0.3,yscale=2.6]
\draw[->] (-10,0) -- (10.3,0);
\draw[->] (0,-0.05) -- (0,1.6);
\draw[-] (-0.3,1.48) -- (0,1.48);
\node at (0,1.72) (a1) {$\theta(t)$};
\node at (11,0) (a2) {$t$};
\node at (-1.2,1.48) (a3) {$\frac\pi2$};
\node at (0,-0.15) (a4) {$t_0$};
\draw[very thick] plot[smooth] coordinates {(-10., 0.0857518)  (-9., 0.0950628)  (-8., 0.106608)  (-7., 
  0.121282)  (-6., 0.140517)  (-5., 0.166737)  (-4., 0.204319)  (-3., 
  0.261799)  (-2., 0.356862)  (-1., 0.523599)  (0., 0.785398)  (1., 
  1.0472)  (2., 1.21393)  (3., 1.309)  (4., 1.36648)  (5., 
  1.40406)  (6., 1.43028)  (7., 1.44951)  (8., 1.46419)  (9., 
  1.47573)  (10., 1.48504)};
\end{tikzpicture}
   \end{center}
  \end{minipage}
  \caption{\small On the left the behaviour of the probabilities in the neighbourhood of a crossover. Generically the levels do not cross and the eigenstates change smoothly from $\ket{\BM_i}$ to $\ket{\BM_j}$, and vice-versa. The behaviour of the mixing angle is shown on the right and goes smoothly from 0 to $\frac\pi2$ through the crossover. Note that if the two states are macroscopically distinct, $|\Delta|\sim \exp[-NX^2/\epsilon^2]$ which is minuscule, and the time scale of the crossover $|p_0\Delta/a|$ is therefore incredibly short: in fact much smaller than the Planck time $10^{-44}\ \text{s}$. Such small time scales will not be resolved in any coarse grained description.}
\label{f4}
\end{figure}
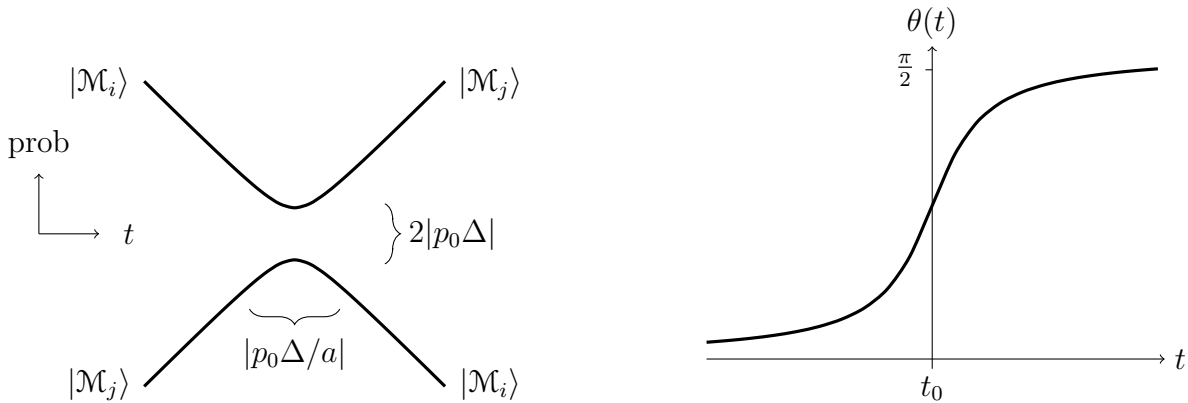

The more generic situation, however, is when $\Delta\neq0$. In this case, the eigenvalues are
\EQ{
p_\pm=p_0\pm\sqrt{(a(t-t_0))^2+|p_0\Delta|^2}
}
and the level crossing is avoided since although the two eigenvalues become close they 
never actually cross. We call this generic situation a crossover as illustrated in figure \ref{f4}. 
It might seem that since we have avoided a degeneracy altogether there is no problem. Unfortunately, this conclusion is too hasty and what actually happens is potentially disastrous and certainly physically unacceptable. To see this we note that near the crossover the eigenvectors are approximately
\EQ{
\ket{\phi_+(t)}&=\cos\theta(t)\ket{\BM_i}+\sin\theta(t)\ket{\BM_j}\ ,\\ \ket{\phi_-(t)}&=-\sin\theta(t)\ket{\BM_i}+\cos\theta(t)\ket{\BM_j}\ ,
}
where we have neglected the time dependence of $\ket{\BM_i}$ and $\ket{\BM_j}$ and for simplicity assumed that $\Delta$ is real. In the above, the angle $\theta(t)$ is given by
\EQ{
\tan\theta(t)=\frac{a(t-t_0)+\sqrt{(a(t-t_0))^2+(p_0\Delta)^2}}{p_0\Delta}\ ,
}
which goes from $0$ to $\frac\pi2$ as $t$ increases through the crossover, as shown in figure \ref{f4}. This reveals the problem: by avoiding the degeneracy the system exhibits an instability in the sense that it switches between the macroscopically distinct states $\ket{\BM_i}$ and $\ket{\BM_j}$ in a very short time of order $|p_0\Delta/a|$. Note that this time scale will be extremely short because the inner product is so small \eqref{ipn}. What this shows is that for an interpretation based on the continuous time process the internal state near a crossover suffers from unacceptably rapid switching between macroscopically distinct states. The existence of crossovers was pointed out in the original work of Zeh \cite{Zeh70,Zeh73, Zeh00,Zeh05}. In that context, Zeh was interested in a many-worlds type interpretation where the crossover it interpreted as 
flip between different branches. In that context the crossover is potentially benign because all branches are supposed to co-exist. In the present context the crossovers present a serious problem for the physical acceptability of the internal state.

\subsection{Coarse graining}\label{s4.3}

The problem with switching instabilities during a crossover arise when one insists on a stochastic process that is continuous in time. In fact the temporal resolution would have to be much smaller than even the Plank time in order to resolve a crossover, a point made in figure \ref{f4}. However, no measuring device could work to this kind of temporal resolution 
and the description with non-relativistic quantum mechanics would in any case break down well before we reached this scale. In order to have a more realistic description we will 
extend our the arguments for coarse gaining in the spatial domain to the temporal domain by introducing a temporal cut off scale $\eta$. The simple way to do this is to define a history at discrete time points $t_n=n\eta$ with $\eta$ small but non-vanishing, as in \eqref{plk}.

If we denote denote the states of the Schmidt decomposition at $t_n$ 
as $\ket{\Psi_i^{(n)}}$ with
\EQ{
\ket{\Psi(t_n)}=\sum_i\sqrt{p_i^{(n)}}\ket{\Psi_i^{(n)}}\ ,\qquad
\langle\Psi_i^{(n)}\ket{\Psi_i^{(n)}}=\delta_{ij}\ ,
}
and define projection operators $\PP_i^{(n)}=\ket{\Psi^{(n)}_i}\bra{\Psi^{(n)}_i}$,
then we have
\EQ{
p_i^{(n+1)}=\sum_jV^{(n)}_{ij}\ ,\qquad p_i^{(n)}=\sum_jV^{(n)}_{ji}\ ,
}
where we have defined
\EQ{
V^{(n)}_{ij}=\RE\,\bra{\Psi(t_{n+1})}\PP^{(n+1)}_iU^{(n)}\PP_j^{(n)}\ket{\Psi(t_n)}\ .
}
Here, $U^{(n)}$ is the unitarity transformation that shifts the state of the complete  system from $t=t_n$ to $t_{n+1}$:
\EQ{
\ket{\Psi(t_{n+1})}=U^{(n)}\ket{\Psi(t_n)}\ .
}
It follows from the above that we can write an expression that is the finite time analogue of \eqref{ge1}
\EQ{
p_i^{(n+1)}=p_i^{(n)}+\sum_jJ_{ij}^{(n)}\ ,\qquad J^{(n)}_{ij}=V^{(n)}_{ij}-V^{(n)}_{ji}\ .
}
It follows that can write a discrete time Markov process \eqref{vv1} defined in analogy with \eqref{trt}, for $i\neq j$, 
\EQ{
p_{ij}^{(n)}=\frac1{p_j^{(n)}}\,\text{max}\left(J_{ij}^{(n)},0\right)\ ,\qquad
p^{(n)}_{ii}=1-\sum_{j\neq i}p_{ji}^{(n)}\ ,
\label{trt2}
}
between time steps $t_n$ and $t_{n+1}$. As in the continuous time process, transitions between pairs of states only go in one direction.

However, there is an important issue that we have hidden in writing the process \eqref{trt2}.
In the coarse graining process there is no a priori relation between the labelling of states at time $t_n$ and $t_{n+1}$. If the $N$ is the dimension of the smallest of the Hilbert sub-spaces $\BH_\BA$ or $\BH_{\BA'}$, there are potentially $N!$ different stochastic processes of the form \eqref{trt2} that one can write down corresponding to the permutations of the labels of the $N$ states at time $t_{n+1}$ relative to those at $t_n$. Not all these processes will be consistent because there is a constraint $p_{ii}^{(n)}\geq0$ but this still leaves an ambiguity. One way to fix this is to rely on continuity in the limit $\eta\to0$ and to fix the labelling of the states by demanding
$\ket{\Psi_i^{(n+1)}}\to\ket{\Psi_i^{(n)}}$. In this limit, the stochastic process \eqref{trt2} reduces to the continuous time process discussed in the last section. However, such a choice runs counter to philosophy of coarse graining and will suffer from the switching problem near crossovers. 

A more natural way to construct a relation between the states and fix the permutation ambiguity is to demand for small but finite $\eta$ (relative to the other time scales in the problem) that
\EQ{
\RE\,\langle\Psi^{(n+1)}_i|U^{(n)}\ket{\Psi^{(n)}_j}\approx\begin{cases}1& i=j\ ,\\
0& i\neq j\ .\end{cases}
\label{jee}
}
This is the unique choice for which 
the transition rates $p_{ij}^{(n)}$, $i\neq j$, are small and $p_{ii}^{(n)}\approx1$. So with
this choice the states have a tendency to endure.

It is important in \eqref{jee} that we do not take the strict $\eta\to0$ limit 
because, as we saw in the last section, when there is a crossover the state $\ket{\Psi^{(n+1)}_i}$ will not converge smoothly to $\ket{\Psi^{(n)}_i}$. In that example and choosing time steps $t_n=t_0+(n+\frac12)\eta$, we find
\EQ{
\Big|\braket{\phi_\pm(t_0-\eta/2)}{\phi_\pm(t_0+\eta/2)}\Big|
&=\frac{|p_0\Delta|}{\sqrt{(a\eta/2)^2+(p_0\Delta)^2}}\ ,\\
\Big|\braket{\phi_\pm(t_0-\eta/2)}{\phi_\mp(t_0+\eta/2)}\Big|
&=\frac{|a\eta/2|}{\sqrt{(a\eta/2)^2+(p_0\Delta)^2}}\ .
}
So keeping $\eta$ small with respect to other time scales in the problem, but still finite, if the crossover has $|p_0\Delta|/a\ll\eta$ the condition \eqref{jee} actually identifies the states as
\EQ{
\ket{\phi_\pm(t_0-\eta/2)}\longleftrightarrow
\ket{\phi_\mp(t_0+\eta/2)}\ .
}
In such an example, the coarse graining procedure means that the crossover is not resolved and effectively appears as a level crossing. Since $|\Delta|\sim\exp[-NX^2/\epsilon^2]$, as in \eqref{ipn}, is extremely small for macroscopically distinct states, the coarse graining in the temporal domain has the effect of curing the unacceptably rapid
switching between macroscopically distinct states that result from the continuous time process.

\section{Decay Processes}\label{s5}

In this section we consider whether the stochastic process described in the last section has any 
observable consequences. In the measurement processes described so far, the 
time dependence of the reduced density matrix over which the collapse occurs 
and the internal states become macroscopically distinct
is a very short time scale $t_\text{c}$ due to efficiency of decoherence. The question is whether are any circumstance when transitions can occur over macroscopic time intervals and have observable effects? It turns out that such transitions do indeed occur in the context of a decay process. Anyone who has listening to a Geiger counter will have first hand experience of such a stochastic process. Our approach follows very closely the analysis of Sudbery \cite{Sud1} but adapted for our purposes.

The set-up involves a decay process which we will think of in terms of a particle decay $A\to B+C$. The Hamiltonian for the system $H=H_0+\lambda H'$
includes a perturbation that is responsible for coupling the initial and final states. The initial state is an eigenstate of the unperturbed Hamiltonian $H_0\ket{A}=E_0\ket{A}$ and the perturbation takes the simple form
\EQ{
H'=\ket{A}\bra{BC}+\ket{BC}\bra{A}\ .
}
Since we expect the decay products to disperse quickly the overlap between the states $\ket{BC}$ and $e^{-iH_0t/\hbar}\ket{BC}$ will have support in only a small region of $t$; that is we will take
\EQ{
\bra{BC}e^{-iH_0t/\hbar}\ket{BC}=0\ ,\qquad |t|>\tau\ .
\label{op1}
}
The approximate solution to the Schr\"odinger equation is given by
\EQ{
\ket{\Psi(t)}=e^{-\gamma t-iE_0t/\hbar}\ket{A}+\frac{\lambda}{i\hbar}\int_0^t dt'\,e^{-\gamma t'-iE_0t'/\hbar}e^{-iH_0(t-t')/\hbar}\ket{BC}\ .
\label{nxx}
}
Here, the integrand at time $t'$ is interpreted as the contribution from the decay of $A$ at the time $t'$ into $B+C$ that then propagates for a time $t-t'$. 
The quantity $(2\gamma)^{-1}$ is lifetime of $A$ determined by
\EQ{
\gamma=\frac{\lambda^2}{\hbar^2}\int_0^\tau dt\,\bra{BC}e^{-iH_0t/\hbar}\ket{BC}\ .
\label{gyq}
}
The validity of the solution \eqref{nxx} requires $\tau\gamma\ll1$.

Now we introduce a measuring device $\BM$ to record the decay products and for it to be realistic we will assume that it has some in-built finite resolution $\eta$ in time, which we assume is much smaller than the lifetime of $A$, $\eta\gamma\ll 1$. The device $\BM$ must have some initial resting state $\ket{\BM_0}$ along with a number of discrete states $\ket{\BM_j(t)}$, $j=1,2,\ldots$, which indicate that $A$ decayed in the time window $t_{j-1}<t<t_j$. In order for $\BM$ to be efficacious, the states $\ket{\BM_j(t)}$, which start equal to the resting state $\ket{\BM_0}$ at $t\sim j\eta$, must, in the ensuing dynamics, become macroscopically distinct states for later times.
The state of the combined system at time $t$ is then\footnote{In the following $[x]$ is the largest integer smaller than $x$.}
\EQ{
\ket{\Psi(t)}=e^{-\gamma t-iE_0t/\hbar}\ket{\BM_0}\otimes\ket{A}+\frac{\lambda}{i\hbar}\sum_{j=1}^{[t/\eta]}\int_{t_{j-1}}^{t_j}dt'\,e^{-\gamma t'-iE_0t'/\hbar}\ket{\BM_j(t)}\otimes e^{-iH_0(t-t')/\hbar}\ket{BC}\ ,
\label{shd}
}
where $t_j=j\eta$.\footnote{It is interesting that if there were a measuring device with infinitely fine temporal resolution then we would find a temporal version of the same problem that was found in section \ref{s2}; namely, the eigenvectors of the reduced density matrix of $\BM$ would be spread out in the temporal domain.} 

In this example, since the measuring device has a finite resolution $\eta$ the temporal coarse graining scale of the stochastic dynamics is set to $\eta$. Choosing time steps $t=t_n$,\footnote{More precisely $t=n\eta-\delta$ with $\delta>0$ and then we take the limit $\delta\to0$ after the limit $\tau\to0$. Alternatively one should take the time steps in the middle of the windows $t_j<t<t_{j+1}$.}
the density matrix of $\BM$ at these discrete points is
\EQ{
\rho(t_0)&=\ket{\BM_0}\bra{\BM_0}\ ,\\
\rho(t_1)&=e^{-2\gamma\eta}\ket{\BM_0}\bra{\BM_0}
+2\gamma\eta\,\ket{\BM_1(t_1)}\bra{\BM_1(t_1)}\ ,\\
\vdots~~~&~~~~~~~~~~~~~~~~~~~~~~~~~~~~~~~\vdots\\
\rho(t_n)&=e^{-2n\gamma\eta}\ket{\BM_0}\bra{\BM_0}+\sum_{j=1}^n 2\gamma\eta\,e^{-2\gamma (j-1)\eta}\ket{\BM_j(t_n)}\bra{\BM_j(t_n)}\ .
}
If the measuring device is to do its job
effectively we require that the 
the state $\ket{\BM_j(t)}$, which at $t=t_j$ is equal to the resting state, becomes a 
distinct macroscopic states after some measurement characteristic time scale $t_\text{c}$ as in \eqref{loc}.
This means that eventually all the states $\ket{\BM_j(t)}$ become
 effectively orthogonal, as in \eqref{ipn}. Therefore for large enough $t$ the eigenvectors of $\rho(t)$ are either $\ket{\BM_0}$, the resting state of $\BM$ indicating that no decay has occurred, with probability $e^{-2\gamma t}$, or one of the $\ket{\BM_j(t)}$, $1\leq j\leq n$, with probability $2\gamma\eta\,e^{-2\gamma (j-1)\eta}$, which indicates that $A$ decayed to $B+C$ in the time window $t_{j-1}<t<t_j$. 

\begin{figure}[ht]
\begin{center}
\begin{tikzpicture}[scale=0.5]
\draw[->] (1,0) -- (11.5,0);
\draw[->] (1,0) -- (1,10.5);
\draw[very thick] (3.5,2.2) -- (11,2.2);
\draw[very thick] (6,1.7) -- (11,1.7);
\draw[very thick] (8.5,1.34) -- (11,1.34);
\draw[-] (3.5,0) -- (3.5,-0.2);
\draw[-] (6,0) -- (6,-0.2);
\draw[-] (8.5,0) -- (8.5,-0.2);
\node at (3.5,-0.8) (a1) {$\eta$};
\node at (6,-0.8) (a1) {$2\eta$};
\node at (8.5,-0.8) (a1) {$3\eta$};
\draw[->,very thick, densely dashed] (1,10) -- (3.5,2.2);
\draw[->,very thick, densely dashed] (3.5,7.8) -- (6,1.7);
\draw[->,very thick, densely dashed] (6,6) -- (8.5,1.34);
\draw[very thick, densely dashed] (8.5,4.7) -- (9.5,3.3);
\node at (12,0) (a1) {$t$};
\node at (0.3,10) (a1) {$1$};
\node at (0.3,0) (a1) {$0$};
\node at (17,4) (b1) {$\ket{\BM_1}$   $p_1=2\gamma\eta$};
\node at (17.9,2) (b2) {$\ket{\BM_2}$   $p_2=2\gamma\eta e^{-2\gamma\eta}$};
\node at (17.9,0) (b3) {$\ket{\BM_3}$   $p_3=2\gamma\eta e^{-4\gamma\eta}$};
\draw[->] (b1) -- (11.1,2.2);
\draw[->] (b2) -- (11.1,1.7);
\draw[->] (b3) -- (11.1,1.34);
\node at (6,10) (c1) {$\ket{\BM_0}$};
\draw[->] (c1) -- (3,8.4);
\node[rotate=90] at (-1,5) (a1) {probability};
\draw[very thick] plot[smooth] coordinates {(1, 10) (2, 9.04) (3, 8.187) (4, 7.4081) (5, 6.703) (6, 6.0653) (7, 
   5.4881) (8, 4.965) (9, 4.4932) (10, 4.065) (11, 3.678)};
\end{tikzpicture}
\end{center}
\caption{\small The behaviour of the probabilities in the decay model with the magnitude of 
$\gamma\eta$ exaggerated for clarity. Transitions occur from the state $\ket{\BM_0}$ to the state $\ket{\BM_j(t)}$ only in the time window $t_{j-1}<t<t_j$ as shown by the dotted lines.}
\label{f7}
\end{figure}
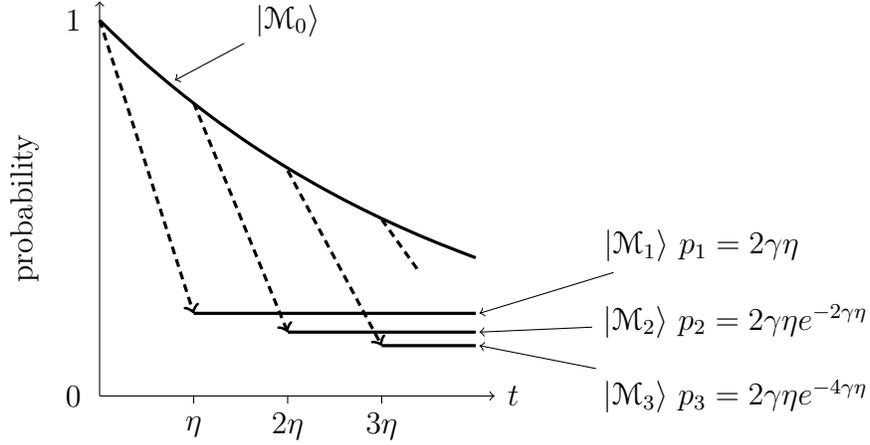

\BOX{{\bf The internal state:}  after a time $T=n\eta$ the internal state is one of the macroscopically distinct states $\ket{\BM_j(t)}$ with probability $p_j=2\gamma\eta\,e^{2\gamma(j-1)\eta}$ indicating that the decay happened in the interval $t_{j-1}<t<t_j$, $j=1,2,\ldots,n$, or remains in the resting state $\ket{\BM_0}$ with probability $p_0=e^{-2n\gamma\eta}$ as illustrated in figure \ref{f7}. This is in accord with the Copenhagen interpretation.}

The probabilities above are exactly what one would expect for a decay process.
However, there is something else that we should expect; namely, that once the system has decayed to $B+C$ it does not then return to the initial state. Just by calculating the single time probabilities as above does not tell us about the actual histories of the internal states and for this we need to use the stochastic process proposed in \ref{s4}. Explicit calculation gives
\EQ{
J^{(n)}_{j0}&=\frac{\lambda^2}{\hbar^2}e^{-\gamma n\eta+iE_0n\eta/\hbar}\int_{t_{j-1}}^{t_j} dt'\,e^{-\gamma t'-iE_0t'/\hbar}\bra{A}H'e^{-iH_0(n\eta-t')/\hbar}\ket{BC}\\ &
=2\eta\gamma e^{-2n\gamma\eta}\delta_{j-1,n}\ ,
}
where we used the fact that $H'\ket{A}=\ket{BC}$ and that
the state of the measuring device $\ket{\BM_j(t)}$, indicating that the particle decayed at time $t_j$, is equal to the resting state $\ket{\BM_0}$ when $t=t_{j-1}$. Of course for subsequent times the two states will quickly become macroscopically distinct and effectively orthogonal. Since $J_{j0}^{(j-1)}>0$ the transition goes from $\ket{\BM_0}\otimes\ket{A}$ to $\ket{\BM_j}\otimes\ket{BC}$, but not the reverse, with conditional probability
\EQ{
p_{j0}^{(n)}=2\eta\gamma\delta_{j-1,n}\ ,
}
as one expects since $(2\gamma)^{-1}$ is the lifetime.
This makes perfect physical sense since we do not expect that the decay products will re-form the original particle. In addition, 
\EQ{
p^{(n)}_{ij}=0\qquad i\neq j\ ,\quad i,j>0\ ,
}
so there are no transitions between the decayed states. In more detail, one finds
\EQ{
\bra{\Psi_0^{(j)}}U^{(j-1)}\ket{\Psi_0^{(j-1)}}=e^{-\gamma\eta}\ ,\qquad
\bra{\Psi_j^{(j)}}U^{(j-1)}\ket{\Psi_0^{(j-1)}}=\sqrt{2\gamma\eta}\ ,
}
which shows how \eqref{jee} works to identify the states in this case. 

It is important to realize in the analysis above that transitions only occur between states that are not yet macroscopically distinct. In this case it is between the states $\ket{\BM_0}$ and $\ket{\BM_j(t_j)}$. It is only for later times that these two states become macroscopically distinct.
In conclusion the stochastic process that we have advocated gives a set of histories that correspond precisely to what a measuring device would record for a decaying system. In particular, the fact that the transitions are only one way is crucial to get an outcome that agrees with our knowledge of decaying systems.

\section{Discussion}

We have attempted to pull together several existing ideas in order to make a precise theory from which the Copenhagen interpretation of quantum mechanics emerges when applied to coarse grained sub-systems with a large number of components. The most important element of the proposal is the distinction between the external and internal views. It is this that allows one to have distinct outcomes and effective collapse for the internal view and linearity for the external view. In a sense the internal view provides a form of local realism but importantly one that does not involve any contradiction with the known quantum violation of Bell's inequalities as shown explicitly by Bene \cite{Bene:1997mf} (see also Smerlak and Rovelli \cite{Smerlak:2006gi}). 

In the interpretation being proposed the relevant internal views are associated to coarse grained subspaces of the full Hilbert space. The degree of coarse graining depends on the resolution of the measuring device being considered, that is the scale at which we are probing the system being measuring. In this respect the resolution scale $\epsilon$ is directly analogous to the Wilsonian cut off in a QFT that can float according to the relevant scale of interest. In that context there is a theory that relates the descriptions with different cut offs known as the renormalization group and it would be interesting to explore the connection with measurements in quantum mechanics in more detail. Some discussion that should be relevant appears in \cite{Balasubramanian:2011wt}.

In this work we have only considered very simple measurement set ups but the results suggest that the classical states are 
closely related to particular coarse grained internal states of macroscopic systems. The picture being suggested here is that the classical world floats like a thick crust on the quantum world underneath and the interface is defined dynamically by the finite resolution of the quasi-classical description.

\appendix
\appendixpage

\section{Exact Models of Continuous Schmidt Eigenstates}\label{a1}

In this appendix, we review two examples considered in the literature where the Schmidt decomposition of 
a continuous system with decoherence 
can be written down exactly. Both cases describe a particle moving in one dimension and an interaction with a second system that leads to an effective reduced density matrix of the form
\EQ{
\rho(x,y)=\psi(x)\psi^*(y)D(x,y)\ ,
}
where the decoherence factor behaves as
$D(x,y)\sim e^{-a(x-y)^2}$ as $x\to y$.

(i) In the first case, discussed by Donald \cite{Donald1}, the particle is confined to the region $x\in[0,L]$. We take the wave function
to be constant $\psi(x)=1/\sqrt{L}$. The decoherence factor needs to be suitably defined 
so that it vanishes for $x,y=0,L$, suggesting
\EQ{
D(x,y)=\frac{\sum_{n=-\infty}^\infty\left[e^{-a(x-y-nL)^2}-e^{-a(x+y-nL)^2}\right]}
{\sum_{n=-\infty}^\infty e^{-a(nL)^2}}\ ,
}
which clearly has the appropriate behaviour as $x\to y$. Then, performing a Poisson re-summation yields the following form for the reduced density matrix
\EQ{
\rho(x,y)=\frac1LD(x,y)=\frac2{ZL}\sum_{n=1}^\infty e^{-(n/2aL)^2}\sin\frac{\pi nx}L\sin\frac{\pi ny}L\ ,
\label{kk2}
}
where $Z=\sum_{n=1}^\infty e^{-(n/2aL)^2}$. 

The form \eqref{kk2} is precisely the eigen-decomposition of $\rho$ with eigenvalues
\EQ{
p_n=\frac{e^{-(n/2aL)^2}}{\sum_{n=1}^\infty e^{-(n/2aL)^2}}\ .
}
The eigenfunctions are simply the stationary states of the infinite square well and are manifestly not localized in space.
In this case, it is worth pointing out that the reduced density matrix has the form of a thermal ensemble of temperature
\EQ{
kT=\frac{2\hbar^2a}{m^2}\ ,
}
in which case $a^{-1/2}$ is the de~Broglie thermal wavelength.

(ii) The second involves a Gaussian wave-packet $\psi(x)=\sqrt{\pi/2b}\,e^{-bx^2}$ and a Gaussian decoherence function
\EQ{
D(x,y)=e^{-a(x-y)^2}\ ,
}
discussed by Bacciagaluppi in \cite{B2} (see also \cite{Barvinsky:1995va}).
The eigenfunctions of the reduced density matrix are given by eigenfunctions of a harmonic oscillator,
\EQ{
\int_{-\infty}^\infty dy\,\rho(x,y)\varphi_n(y)=p_n\varphi_n(x)\ ,
}
with
\EQ{
\varphi_n(x)=N_nH_n(\sqrt{2\mu}\, x)e^{-\mu x^2}\ ,\qquad p_n=p_0(1-p_0)^n\ ,
\label{hee}
}
$n=0,1,2,\ldots$, and
\EQ{
\mu^2=b(b+2a)\ ,\qquad p_0=\frac{2\sqrt{b}}{\sqrt{b}+\sqrt{2a+b}}\ .
}
In the above, $H_n(x)$ are the Hermite polynomials and $N_n$ is the usual normalizing factor of the harmonic oscillator.

The Gaussian factor in \eqref{hee} suggest that the eigenfunctions are localized over length scale $\mu^{-1/2}$; however, this is too hasty because, as $n$ increases the polynomial pre-factors have the effect of spreading the wave function further and for large $a$ the probabilities $p_n$ are significant even for large $n$. We can estimate the spreading by calculating the average over the ensemble. Given that $(\Delta x_n)^2\sim n/\mu$, we have
\EQ{
(\Delta x)^2=\sum_{n=0}^\infty p_n(\Delta x_n)^2\thicksim\frac{1-p_0}{\mu p_0}\thicksim b^{-1}\ ,
}
for large $a$. In other words, the eigenfunctions of the reduced density matrix are on average no more localized than the original wave-packet.

\section{A Simple Environment}\label{a3}

In this appendix we investigate a crude model of an environment $\BE$ coupled to a 
coarse grained particle. In this model
$\BE$ is composed of a collection of $N$ spin $\frac12$ systems each coupled separately to the particle. For the particle, we first introduce a coarse grained position operator
\EQ{
\tilde x\ket{x}=x_j\ket{x}\ ,\qquad\text{if}\quad x_j<x<x_{j+1}\ .
}
The fact that $\tilde x$ is only an approximate position operator models the fact that we have a coarse grained description of the particle.
In order to draw conclusions we will need $N$ to be macroscopically large so that the particle initiates the movement of a large number of degrees of freedom. 
In the simplest case we shall ignore any interactions between the spins forming $\BE$. This is not very realistic but it sufficient to exhibit decoherence. Let us take the initial state of $\BE$ to consist of all the spins pointing along the $z$-axis $\ket{\phi_0}=\ket{z^+}\otimes\cdots\otimes\ket{z^+}$ (there would be no change in conclusion if the spins were taken to point in random directions).\footnote{$\sigma_{x,y,z}$ are the usual spin operators and $\sigma_z\ket{z^\pm}=\pm\ket{z^\pm}$ are the spin up and down eigenstates in the $z$ direction.} We will choose a simple interaction of the form
\EQ{
H=\sum_{a=1}^Ng_a\sigma_y^{(a)}\otimes\tilde x\ ,
}
where $g_a$ are a set of individual couplings. In this case, we can write the state of the total system as \eqref{ww1}
\EQ{
\ket{\Psi(t)}=\sum_j\ket{\BE_j(t)}\otimes\int_{x_j}^{x_{j+1}} dx\,\psi(x)\ket{x}\ .
}
and solve the Schr\"odingier equation. Note that acting on the component $\ket{\BE_j(t)}$, the Hamiltonian has the effect of 
rotating the $a^\text{th}$ spin by an angle $2g_a x_jt$ along the $y$-axis so that 
\EQ{
\ket{\BE_j(t)}=\bigotimes_{a=1}^N\Big(\cos(g_ax_jt)\ket{z^+}+\sin(g_ax_jt)\ket{z^-}\Big)\ .
}
In this model, the inner product
\EQ{
\langle\BE_i(t)\ket{\BE_j(t)}=\prod_{a=1}^N\cos(2g_a(x_i-x_j)t)\ .
}
As long as the couplings $\{g_a\}$ are suitably spread and when $N$ is very large this may be approximated by a simple Gaussian: 
\EQ{
\langle\BE_i(t)\ket{\BE_j(t)}\thicksim \exp\big[-N\lambda^2t^2(x_i-x_j)^2\big]=\exp\big[-N(\lambda\epsilon t)^2(i-j)^2\big]\ ,
}
where $\lambda^2=2N^{-1}\sum_ag_a^2$. The reason for having different couplings $\{g_a\}$ is that it avoids the problem of recurrences. Notice how the decoherence time $N^{-1/2}(\lambda\epsilon)^{-1}$ scales with $N$ and so will be very short when $\BE$ is macroscopically large. 

\section{Imperfect Measurements}\label{a2}

In this appendix we consider the measurement process when $\BM$ is not 100\% efficient in the sense that there is a probability that the result recorded by $\BM$ is incorrect. The fact that this could happen was used by Albert and Loewer \cite{AL1} to show that using the Schmidt decomposition leads to states that involve macroscopic superpositions. Bacciagaluppi and Hemmo \cite{BH1} subsequently argued that this physically unacceptable conclusion is avoided if the decohering effects of the environment are taken into account. Here, we will consider the role of the environment in our own 
context and also derive an expression for the resulting probabilities.

Firstly, if we do not take account of the environment and when the measuring device is not 100\% efficient, then  the state of the combined system
$\BM+\BP$ for large enough $t$ is expected to be of the form
\EQ{
\ket{\Psi(t)}=\sum_{ij=1}^n\sqrt{p_j}Z_{ij}\ket{\BM_i(t)}\otimes\ket{\psi_j(x)}\ .
\label{vv6}
}
Here, $Z_{ij}$ are the components of a matrix whose off-diagonal terms describe the imperfections of $\BM$ where result $\ket{\BM_i}$ can be indicated even when the particle is in the region $x_j<x<x_{j+1}$. The problem is that when $Z_{ij}$ has off-diagonal terms the eigenvectors of the reduced density matrix involve linear combinations of the macroscopic states $\ket{\BM_j(t)}$ of the approximate form
\EQ{
\sum_{i=1}^nZ_{ij}\ket{\BM_i(t)}\ .
}
This is a potential disaster for our interpretation.
However, when we include the environment in the description the problem is resolved. To see this, one could write the total state including the environment as
\EQ{
\ket{\Psi(t)}=\sum_{ij=1}^n\sum_{a=1}^{m_j}\sqrt{p_j}Z_{aij}\ket{\BM_{ai}(t)}\otimes\ket{\psi_j(x)}\otimes\ket{\BE_{aij}(t)}\ ,
\label{vv9}
}
where the environment states $\ket{\BE_{aij}(t)}$ are expected to become approximately orthogonal in all their labels for large $t$. The fact that we must trace over the environment insures that the 
the reduced density matrix of $\BM$ is now approximately block diagonal: $\rho=\sum_j\rho^{(j)}$. Each block corresponds to a set of states $\ket{\BM_{aj}}$ that are tightly clustered around
a particular macroscopic state indicating that a particle was detected in the region $x_j<x<x_{j+1}$. But now the probability in each block is not equal to $p_j$ due to the imperfections of $\BM$. Rather the total probability associated to the $j^\text{th}$ block is
\EQ{
\text{Tr}\,\rho^{(j)}=\sum_{aik}Z_{aji}Z_{ajk}^*\sqrt{p_ip_k}\neq p_j\ .
}
Note that the total probability sums correctly to one since $\ket{\Psi(t)}$ in \eqref{vv9} is normalized. Importantly, the eigenvectors now are approximately linear combinations $\ket{\BM_{aj}(t)}$ for fixed $j$ and so do not involve linear combinations of macroscopically distinct states. So imperfections of the measuring device do not lead to macroscopic superpositions of $\BM$. On the contrary, if $\BM$ registers an outcome $i$, that is in one of the states $\ket{\BM_{ai}(t)}$,  then the particle-environment system is in the corresponding internal state
\EQ{
\sum_{j=1}^n\sqrt{p_j}Z_{aij}\ket{\psi_j(x)}\otimes \ket{\BE_{aij}(t)}\ .
}

\end{document}